# A Contribution to Theory of Factor Income Distribution, Cambridge Capital Controversy and Equity Premium Puzzle

*By* Xiaofeng Liu[*]

## Abstract

Under very general conditions, we construct a micro-macro model for closed economy with a large number of heterogeneous agents. By introducing both financial capital (i.e. valued capital——equities of firms) and physical capital (i.e. capital goods), our framework gives a logically consistent, complete factor income distribution theory with micro-foundation. The model shows factor incomes obey different distribution rules at the micro and macro levels, while marginal distribution theory and no-arbitrage principle are unified into a common framework. Our efforts solve the main problems of Cambridge capital controversy, and reasonably explain the equity premium puzzle. Strong empirical evidences support our results.

[*] *School of Finance, Nankai University.* E-mail: lxfnku@nankai.edu.cn

# 1. Introduction

Since the 1970s, gradually blurred has been the difference between analytical methods in macroeconomics and microeconomics. For instance, by introduction of the representative agent framework, the proposition——profit and wage are respectively determined by the marginal contributions of capital and labor——is directly extended from the micro level to the macro level. However, in a famous paper——"More is Different," P.W. Anderson (1972), the Nobel laureate in Physics, clarified and highlighted the principle——understanding the characteristics of the individuals and the laws at the micro level does not mean that we have grasped the nature of macro-system which contains a large number of micro-agents. In fact, the macro-characteristics are often "emerged" with the increase in the quantity of them. Unfortunately, it is essentially violated by the popular representative agent framework in macroeconomics.[1]

In our view, the mainstream approach has led to a series of paradoxes and problems. At the theoretical level, they are mainly related to the Cambridge capital controversy, such as the capital-aggregation problem, reswitching and capital reversing, Pasinetti paradox, intrinsic inconsistency between general equilibrium framework and no-arbitrage principle, etc. In empirical research, at least, it includes the equity premium puzzle.[2] All of these challenges are related to the incompleteness of the theories of factor income distribution. Establishment of a new micro-macro framework that simultaneously solves all of these problems, will be the cornerstone of an intrinsically consistent and complete value theory.

---

[1] The representative agent hypothesis lacks logical basis (see Kirman(1992, 2009), J.E. Stiglitz(2019), etc.).

[2] As long-term unsolved problems and paradoxes, the related researches involve a large amount of literature. Due to space limitations, we just list a few excellent review articles. About the Cambridge capital controversy, see Cohen and Harcourt (2003), Pasinetti (2003), Lazzarini (2011), Garegnani (2012), etc.; the equity premium puzzle, see Mehra R. (2003), etc.

Inspired by the above ideas, by constructing the model with a large number of heterogeneous households, heterogeneous firms, and heterogeneous products, this paper gives the answers to all of the above problems. On the assumption side, our model only requires: (i) there are enough heterogeneous households and heterogeneous firms; (ii) firms distribute dividends according to the number of shares. The linchpin to our analysis is the introduction of two different concepts of capital (valued capital and capital goods, i.e. financial capital and physical capital). Here, valued capital is the market value of firm's equities. Correspondingly, capital goods is a physical set of various machines and production materials. Our main results include: (i) at the micro level, the values of firms' equities are determined by the no-arbitrage principle, while general equilibrium determines relative prices of capital goods; (ii) at the macro level, the rate of return on market portfolio is related to the saving rates of households, and in an asymptotic sense, determined by the saving rate of the richest group of households in the economy. These results mean that profit rates based on valued capitals are independent of the marginal principle. Our contribution solves the main problems related the Cambridge capital controversy and reasonably explains the equity premium puzzle. The new framework provides an alternative approach to construct a more rational micro-macro model.

Following this introduction, Section 2 introduces the model. Section 3 discusses the theoretical implications of the model and the solutions to the related problems. Section 4 provides empirical evidences. Section 5 draws the conclusions.

## 2. Model

### 2.1. *A Closed Economy*

Let us consider a closed economy with an external observer, Alice. In this paper, we require that Alice's observations are all ex-post facto. Correspondingly, our analysis below focuses on the laws behind these ex-post data. Respectively, let $\mathcal{H}$, $\mathcal{Q}$, and $\mathcal{F}$ be the sets of all possible households, all possible products, and all possible firms, in all periods. They are defined by

$$\mathcal{H} \equiv \left\{ h_i \mid i \in \mathbb{Z}^+ \right\}, \quad \mathcal{Q} \equiv \left\{ q_i \mid i \in \mathbb{Z}^+ \right\}, \mathcal{F} \equiv \left\{ f_i \mid i \in \mathbb{Z}^+ \right\}.$$

All households in period $t$, $H(t)$ ($H(t) \subset \mathcal{H}$, $\forall t$), own all firms $F(t)$, ($F(t) \subset \mathcal{F}$, $\forall t$). At the beginning of period $t$, $Q(t)$ is the set of products, ($Q(t) \subset \mathcal{Q}$, $\forall t$).[3] $Q(t)$ contains both capital goods and consumer goods.

#### (*) *Firms*

In period $t$, the stock of capital goods $\mathbf{K}^t$ is defined by

$$\mathbf{K}^t = \left( k_{i,j}^t \right)_{|F(t)| \times |Q(t)|}, \ k_{i,j}^t \geq 0, \ \forall i, \ \forall j,$$

where $k_{i,j}^t$ is the quantity of $q_j (q_j \in Q(t))$, which is owned by the firm $f_i$ ($f_i \in F(t)$).

It must take at least one period for any meaningful production activity, and will end in a limited period of time——$\mathcal{T}$ is the maximum number of periods, $\mathcal{T} \in Z^+$. $S_i^t$, the technique of $f_i$ during the period $t$, is defined as a mapping from the input-set $X(t)$ to the output-set $Y(t+1)$, i.e.

---

[3]For the convenience of introduction, we first temporarily set $H(t)$, $F(t)$ and $Q(t)$ to be unchanged. In subsection 3.5, we demonstrate that their time variations have no effect on our results.

$$S_i^t: \ X(t) \to Y(t+1),$$

$$X(t) \equiv \mathbb{R}^{|Q(t)|} \times \mathbb{R}^{|H(t)|} \times ... \times \mathbb{R}^{|Q(t+\mathcal{T}-1)|} \times \mathbb{R}^{|H(t+\mathcal{T}-1)|},$$

$$Y(t+1) \equiv \mathbb{R}^{|Q(t+1)|} \times \mathbb{R}^{|Q(t+2)|} \times ... \times \mathbb{R}^{|Q(t+\mathcal{T})|}.$$

Under the above definition, respectively, $X(t)$ and $Y(t+1)$ are the sets of all possible inputs and outputs that are distributed at different periods along the time axis. When $\mathcal{T}=1$, $S_i^t$ will be reduced to a traditional production function, i.e.

$$S_i^t: \ X(t) \to Y(t+1), \ \ X(t) = \mathbb{R}^{|Q(t)|} \times \mathbb{R}^{|H(t)|}, \ Y(t+1) = \mathbb{R}^{|Q(t+1)|}.$$

In this case, respectively, a specific input $x^t$ and a specific $y^{t+1}$ satisfy

$$x^t = \left([x_1^t, x_2^t, ..., x_{|Q(t)|}^t]^T, [l_1^t, l_2^t, ..., l_{|H(t)|}^t]^T\right), \ \ x^t \in X(t);$$

$$y^{t+1} = \left([y_1^{t+1}, y_2^{t+1}, ..., y_{|Q(t+1)|}^{t+1}]^T\right), \ \ y^{t+1} \in Y(t+1),$$

where, respectively, $x_i^t$, $l_i^t$ and $y_j^{t+1}$ represent the quantity of the product $q_i (q_i \in Q(t))$, the working hours of the households $h_i (h_i \in H(t))$, and the quantity of the product $q_j (q_j \in Q(t+1))$.

In the real economy, all physical capitals will depreciate over time. For this situation, we can technically treat the capital goods that have been depreciated in period $t$ as output, and record them into $Y(t+1)$.

In period $t$, the values of firms ($\mathbf{A}^t$) are defined as

$$\mathbf{A}^t = \left[A_1^t, A_2^t, ..., A_{|F(t)|}^t\right]^T,$$

where $A_i^t$ is the market value of the equities of the firm $f_i$ ($f_i \in F(t)$) at the beginning of the period $t$. Generally, unless by chance, $A_i^t$ does not equal the total market value of all capital goods of $f_i$, even if the prepaid wages are included. In other words, we almost always have

$$\mathbf{A}^t \neq \mathbf{K}^t \mathbf{P}^t, \quad \mathbf{A}^t \neq \mathbf{K}^t \mathbf{P}^t + diag\left(\mathbf{L}^t \mathbf{W}^t\right),$$

where the elements of $\mathbf{P}^t$ are commodity prices, and defined as

$$\mathbf{P}^t = [p_1^t, p_2^t, ..., p_{|Q(t)|}^t]^T, \quad p_i^t \geq 0, \forall i.$$

Here, $p_i^t$ represents the nominal price of the product $q_i (q_i \in Q(t))$ in period $t$. Respectively, $\mathbf{L}^t$ and $\mathbf{W}^t$ represent the working hours and the nominal wages of households, i.e.

$$\mathbf{L}^t = \left(l_{i,j}^t\right)_{|F(t)| \times |H(t)|}, \quad \mathbf{W}^t = \left(w_{i,j}^t\right)_{|H(t)| \times |F(t)|}, \quad \sum_{i=1}^{|F(t)|} l_{i,j}^t \leq 1,$$

where $l_{i,j}^t$ is the labor hours of household $h_j (h_j \in H(t))$ in period $t$, that is employed by firm $f_i (f_j \in F(t))$. $w_{j,i}^t$ is the nominal wage that firm $f_i$ pays household $h_j$ in period $t$.

Relative to the physical production process determined by techniques, in the sense of funds, cash flows of firm $f_i (f_j \in F(t))$ are affected by many factors, including payment habits (prepayment or post-payment), technique ($S_i^t, t = 1, 2, ...$) and prices ($\mathbf{P}^t, t = 1, 2, ...$), etc. Let $M(t)$ be the set of all possible net cash flows faced by $f_i$, and satisfy

$$\mathbf{m}_i^t = [m_i^t, m_i^{t+1}, ..., m_i^{t+\mathcal{T}}]^T, \quad \mathbf{m}_i^t \in M(t), \quad M(t) \subset \mathbb{R}^{\mathcal{T}+1},$$

where $m_i^{t+j}$ is net cash income of $f_i$ in period $t+j$.

From an after-the-fact point of view, the decision of firm $f_i$ is essentially equivalent to select a $\mathbf{m}_i^t$ from $M(t)$. Simultaneously, along with the production process, $A_i^t$ can change over time. At the end of period $t$, the rate of return for the shareholders of firm $f_i$——$z_i^{t+1}$ satisfies

$$z_i^{t+1} = (A_i^{t+1} - A_i^t) / A_i^t, \quad \forall f_i \in F(t). \tag{1}$$

From (1), we can define

$$\mathbf{Z}^{t+1} \equiv \left[ z_1^{t+1}, z_2^{t+1}, ..., z_{|F(t)|}^{t+1} \right]^T.$$

In period $t$, $\mathbf{Z}^{t+1}$ is influenced by a variety of factors, including expectations, financing methods, payment habits, etc., not just by techniques $S_i^t$ ($\forall f_i \in F(t)$).

### (*) *Households*

The labor incomes ($\hat{\mathbf{W}}^t$) of households in period $t$ are defined by

$$\hat{\mathbf{W}}^t = \left[ \hat{w}_1^t, \hat{w}_2^t, ..., \hat{w}_{|H(t)|}^t \right]^T = diag\left( \mathbf{W}^t \times \mathbf{L}^t \right),$$

where $\hat{w}_i^t$ is the labor income of household $h_i$ ($h_i \in H(t)$) in period $t$. $\mathbf{Q}^t$ is defined by

$$\mathbf{Q}^t = \left( q_{i,j}^t \right)_{|H(t)| \times |Q(t)|},$$

where $q_{i,j}^t$ is the quantity of product $q_j$ ($q_j \in Q(t)$) purchased by household $h_i$. In principle, $\mathbf{Q}^t$ can be calculated by general equilibrium in a competitive economy. However, we do not intend (and do not need) to discuss the decision-making process of households. So, in the analysis below, $\mathbf{Q}^t$ is seen as exogenous.

Respectively, in period $t$, nominal consumption expenditures of households $\hat{\mathbf{C}}^t$, and nominal net assets of households $\hat{\mathbf{A}}^t$ satisfy

$$\hat{\mathbf{C}}^t = \mathbf{Q}^t \mathbf{P}^t = \left[c_1^t, c_2^t, ..., c_{|H(t)|}^t\right]^T, \quad \hat{\mathbf{A}}^t = \left[a_1^t, a_2^t, ..., a_{|H(t)|}^t\right]^T,$$

where, respectively, $c_i^t$, $a_i^t$ are nominal consumption and nominal net asset value of household $h_i$.

Investment return rates of households are defined by $\mathbf{r}^{t+1}$, i.e.

$$\mathbf{r}^{t+1} = \left[r_1^{t+1}, r_2^{t+1}, ..., r_{|H(t)|}^{t+1}\right]^T,$$

where $r_i^{t+1}$ is the rate of return on the portfolio of $h_i$. We define $\mathbf{V}^t$ as

$$\mathbf{V}^t \equiv \left(\omega_{i,j}^t\right)_{|H(t)| \times |F(t)|},$$

where $\omega_{i,j}^t$ is the weight of the portfolio of household $h_i$ on the firm $f_j$ in period $t$. So, we have

$$\mathbf{r}^{t+1} = \mathbf{V}^t \mathbf{Z}^{t+1}. \tag{2}$$

Let $\mathbf{I}_{|H(t)| \times 1} \equiv [1,1,...1]^T$, and $d_i^t$ be nominal debt of household $h_i$ ($h_i \in H(t)$). Respectively, $\hat{\mathbf{D}}^t$ and exogenous $\hat{\mathbf{r}}^{t+1}$ are defined by

$$\hat{\mathbf{D}}^t = \left[d_1^t, d_2^t, ..., d_{|H(t)|}^t\right]^T, \qquad \hat{\mathbf{r}}^{t+1} = \overline{r}_{t+1} \mathbf{I},$$

where $\overline{r}_{t+1}$ is the exogenous interest rate of loans from period $t$ to $t+1$.

The budget constraints that households need to meet are

$$\hat{\mathbf{C}}^t + \hat{\mathbf{A}}^{t+1} - \hat{\mathbf{A}}^t = \hat{\mathbf{A}}^t \circ \mathbf{r}^{t+1} + \hat{\mathbf{W}}^t + \hat{\mathbf{D}}^t \circ \left(\mathbf{r}^{t+1} - \hat{\mathbf{r}}^{t+1}\right), \tag{3}$$

where the operator "$\circ$" means "Hadamard product." Saving rates of households are seen as exogenous, and defined by $\hat{\mathbf{s}}^t$, i.e.

$$\hat{\mathbf{s}}^t = \left[s_1^t, s_2^t, ..., s_{|H(t)|}^t\right]^T,$$

where $s_i^t$ is the saving rate of household $h_i$ in period $t$. From (3), we have

$$\hat{\mathbf{A}}^{t+1} - \hat{\mathbf{A}}^t = \hat{s}^t \circ \left[\hat{\mathbf{A}}^t \circ \mathbf{r}^{t+1} + \hat{\mathbf{W}}^t + \hat{\mathbf{D}}^t \circ \left(\mathbf{r}^{t+1} - \hat{\mathbf{r}}^{t+1}\right)\right]. \tag{4}$$

**(\*) *Debt***

In every period, households take loan by mortgaging their assets. For simplicity, we assume the amount of loans to households only depend on the quantities of mortgage assets that households possess. Here, the mortgage assets are respective investment portfolios of households. So, we have

$$d_i^t = \mu_i^t a_i^t, \quad \forall h_i, h_i \in H(t), \tag{5}$$

where $\mu_i^t$ is exogenous, and related to the specific portfolios. For $\forall i$, $\forall t$, we specify $0 < \mu_i^t < 1$, and define

$$\mathbf{M}^t = \left[\mu_1^t, \mu_2^t, ..., \mu_{|H(t)|}^t\right]^T.$$

Thus, we have

$$\hat{\mathbf{D}}^t = \mathbf{M}^t \circ \hat{\mathbf{A}}^t. \tag{6}$$

### 2.2. *Aggregation, Relative Prices*

At the end of each period, Alice can calculate total consumption $C_t$, total net assets $A_t$, total labor incomes $W_t$ and total debt $D_t$. Respectively, they are defined as

$$C_t \equiv \mathbf{I}^T \hat{\mathbf{C}}^t, \quad A_t \equiv \mathbf{I}^T \hat{\mathbf{A}}^t, \quad W_t \equiv \mathbf{I}^T \hat{\mathbf{W}}^t, \quad D_t \equiv \mathbf{I}^T \hat{\mathbf{D}}^t. \tag{7}$$

Pre-multiplying both sides of (3) by $\mathbf{I}^T$ gives

$$C_t + A_{t+1} - A_t = A_t R_{t+1} + W_t + D_t(R_{t+1} - \bar{r}_{t+1}), \tag{8}$$

where $R_{t+1}$ is the rate of return on market portfolio from period $t$ to $t+1$. The saving rate of economy $s_t$ satisfies

$$s_t\left(C_t + A_{t+1} - A_t\right) = \mathbf{I}^T\left\{\hat{\mathbf{s}}^t \circ \left[\hat{\mathbf{A}}^t \circ \mathbf{r}^{t+1} + \hat{\mathbf{W}}^t + \hat{\mathbf{D}}^t \circ \left(\mathbf{r}^{t+1} - \hat{\mathbf{r}}^{t+1}\right)\right]\right\}, \tag{9}$$

$$A_{t+1} - A_t = s_t\left[A_t R_{t+1} + W_t + D_t(R_{t+1} - \bar{r}_{t+1})\right]. \tag{10}$$

In period $t$, for total net assets $A_t$ and total debt $D_t$, we have

$$D_t = \mu_t A_t. \tag{11}$$

At the same time, $\mu_t$ satisfies

$$\mu_t A_t = \left(\mathbf{M}^t\right)^T \hat{\mathbf{A}}^t. \tag{12}$$

Let $\mathbf{J}_{|F(t)|\times 1} \equiv [1,1,...1]^T$. According to the definitions, we have

$$A_t \equiv \mathbf{J}^T \mathbf{A}^t, \qquad R_{t+1} = (1/A_t)\left[\mathbf{J}^T\left(\mathbf{A}^t \circ \mathbf{Z}^{t+1}\right)\right]. \tag{13}$$

From (13) and (1), we get

$$A_{t+1} = A_t(1 + R_{t+1}). \tag{14}$$

At the micro level, the relative prices (including relative wages) can be determined by "technical relationship" (i.e., preferences, input-output relationship or production function). If the economy is perfectly competitive, the relative prices can be calculated by general equilibrium, although it may be difficult or even impossible in an imperfect competition economy. However, for our model, as long as there exists a relative price system determined by some exogenous factors, our analysis below will not be affect-ed. Without loss of generality, let $p_x^t$ be the nominal price of a product $q_x$ ($q_x \in Q(t)$). Obviously, $\hat{\mathbf{W}}^t$ satisfies

$$\hat{\mathbf{W}}^t = p_x^t \left( \hat{\mathbf{W}}^t / p_x^t \right), \tag{15}$$

where the relative labor incomes $\hat{\mathbf{W}}^t / p_x^t$ are determined by exogenous "technical relationship." Simultaneously, nominal consumption expenditures of households ($\hat{\mathbf{C}}^t$) satisfy

$$\hat{\mathbf{C}}^t = p_x^t \left[ \mathbf{Q}^t \left( \mathbf{P}^t / p_x^t \right) \right]. \tag{16}$$

Similarly, both $\mathbf{Q}^t$ and the relative prices $\mathbf{P}^t / p_x^t$ are determined by exogenous "technical relationship."

So far, the exogenous variables include: (i) $\hat{\mathbf{W}}^t / p_x^t$, $\mathbf{P}^t / p_x^t$ and $\mathbf{Q}^t$, determined by "technical relationship;" (ii) $\hat{\mathbf{s}}^t$ and $\mathbf{V}^t$, determined by heterogeneous households at the micro level; (iii) $\mathbf{M}^t$, related to the portfolios of households and financial market; (iv) $\bar{r}_{t+1}$, determined by demand and supply in money market. However, we have not given the determination equation of $\mathbf{Z}^{t+1}$, so $\mathbf{r}^{t+1}$ is still undetermined. In fact, if no more conditions are added, the above equation system will be indefinite. Fortunately, based on the no-arbitrage principle, under the condition that there exist enough heterogeneous households and heterogeneous firms in economy (i.e., $|H(t)| \mapsto +\infty, |F(t)| \mapsto +\infty$), the problem can be solved.

### 2.3. *No Arbitrage Principle and System Dynamics*

#### (\*) *Rates of Returns on Equities of Firms*

In period $t$, $\mathbf{r}^{t+1}$ is determined by $\mathbf{z}^{t+1}$ (see (2)). However, $\mathbf{z}^{t+1}$ is affected by many factors, including expectations, market structure, technical shocks, etc. So, it is difficult to give a generalized formula for $\mathbf{z}^{t+1}$. Fortunately, according to the results of Ross (1976), when $|H(t)| \mapsto +\infty$, $|F(t)| \mapsto +\infty$, and economy satisfies the no-arbitrage principle, we have[4]

$$\mathbf{z}^{t+1} \approx \overline{\mathbf{r}}^{t+1} + \mathbf{B}^t (R_{t+1} - \overline{r}_{t+1}), \tag{17}$$

where $\overline{\mathbf{r}}^{t+1} = \overline{r}_{t+1} \hat{\mathbf{I}}$, $\hat{\mathbf{I}}_{|F(t)| \times 1} \equiv [1,1,...,1]^T$. The exogenous $\mathbf{B}^t$ is defined by

$$\mathbf{B}^t \equiv \left[ \beta_1^t, \beta_2^t, ..., \beta_{|F(t)|}^t \right]^T, \quad \forall f_j \in F(t).$$

In (17), the symbol "$\approx$" means, for $\forall \varepsilon > 0$, there exists an upper bound on the number of the asset $j$, while the asset $j$ satisfies

$$\left| z_j^{t+1} - \left[ \overline{r}_{t+1} + \beta_j^t \left( R_{t+1} - \overline{r}_{t+1} \right) \right] \right| \geq \varepsilon, \quad f_j \in F(t).$$

In other words, when $|F(t)| \mapsto +\infty$, most assets satisfy (17) in an approximate sense. Essentially, (17) can be seen as a modern version of the unified-rate-of-return theory in classical economics.

(*) *Determination of Endogenous Variables*

Under the above conditions ($|H(t)| \mapsto +\infty, |F(t)| \mapsto +\infty$), all endogenous variables in our model can be determined. At the micro level, these variables include: $\mathbf{z}^{t+1}$, $\mathbf{r}^{t+1}$, $\hat{\mathbf{C}}^t$, $\hat{\mathbf{A}}^t$, $\mathbf{A}^t$, $\hat{\mathbf{W}}^t$, $\hat{\mathbf{D}}^t$, $p_x^t$. At the macro level, they are: $C_t$, $A_t$, $W_t$, $D_t$, $R_{t+1}$. Correspondingly, the exogenous variables include: $\hat{\mathbf{r}}^{t+1}$, $\overline{\mathbf{r}}^{t+1}$, $\overline{r}_{t+1}$, $\mathbf{B}^t$, $\mathbf{V}^t$, $\mathbf{M}^t$, $\hat{\mathbf{W}}^t / p_x^t$, $\mathbf{Q}^t$, $\mathbf{P}^t / p_x^t$.

---

[4] For simplicity, the pricing factor only includes rate of return on market portfolio.

The system $E_1$ consists of the equations: (1), (2), (3), (4), (6), (7), (13), (14), (15), (16), (17). Giving the values of the exogenous variables and the initial values of all variables, the dynamic of economy can be calculated by system $E_1$.

## 2.4. *Asymptotic Nature of the System*

The system $E_1$ can completely determine factor income distribution at both the micro level and the macro level. Generally, due to complexity, it is difficult for us to understand the meaning behind the dynamic of the system $E_1$. However, when $t \to +\infty$, the asymptotic nature of $E_1$ can show some interesting results, and deepen our insight into the dynamics of the economy and factor income distribution theory.

### (*) *Partition of Households*

As can be seen from (9), in general, the saving rate of economy $s_t$ is time-varying, which is affected by income and wealth distributions, even if every element in $\hat{s}^t$ remains constant. Obviously, it leads to analytical difficulties.

An apparent fact is that in a closed economy, all households as a whole ($H(t)$) must hold market portfolio in every period (see equation (8)). So, when the number of households is large enough, we can always get some subsets from $H(t)$, and the portfolios of the households in these subsets are enough to approximate market portfolio. Thus, we get Proposition 1.

Proposition 1: Let $\mathbf{V}_{M(t)}$ be market portfolio. For $\forall \varepsilon > 0$ and $\forall \bar{M} \geq 2$, $\exists \bar{N} > 0$, we have: when $|H(t)| \vartriangleright \bar{N}$, $\exists \mathscr{K}(t)$, where $\mathscr{K}(t)$ is a partition of $H(t)$, $\mathcal{K}(t) \equiv \{H_i(t) \mid i = 1,2,...,\bar{M}\}$, and for $\forall i$, $\mathbf{V}_{H_i(t)}$, which is the portfolio of the households in $H_i(t)$, satisfies

$$\text{dist}\left(\mathbf{V}_{H_i(t)}, \mathbf{V}_{M(t)}\right) = \sqrt{\sum_{j \in F(t)} \left(\omega^t_{Hi,j} - \omega^t_{m,j}\right)^2} < \varepsilon,$$

and $H(t) = \bigcup_{\forall H_i(t) \in \mathcal{K}(t)} H_i(t)$; $\varnothing = H_i(t) \cap H_j(t), \forall H_i(t), H_j(t) \in \mathcal{K}(t), i \neq j$,

$$\mathbf{V}_{H_i(t)} \equiv \left(\omega^t_{Hi,j}\right)_{|F(t)| \times 1}, \quad \mathbf{V}_{M(t)} \equiv \left(\omega^t_{m,j}\right)_{|F(t)| \times 1},$$

where the operator "dist" means "Euclidean distance," and $\omega_{Hi,j}$, $\omega_{m,j}$ are the weights of asset $j$ in portfolios $\mathbf{V}_{H_i(t)}$ and $\mathbf{V}_{M(t)}$, respectively.

Without loss of generality, we assume that $s_{H_1(t)} > s_{H_j(t)}$, $(\forall H_j(t) \in \mathcal{K}(t), j \neq 1, \forall t > 1)$. Here, $s_{H_k(t)}$ is the saving rate of the households in $H_k(t)$, $(\forall H_k(t) \in \mathcal{K}(t), \forall t > 1)$, i.e.

$$s_{H_k(t)} = \left\{\sum_{\forall h_i \in H_k(t)} s_i^t \left[a_i^t r_i^{t+1} + w_i^t + d_i^t(r_i^{t+1} - \bar{r}_{t+1})\right]\right\} \bigg/ \left[\sum_{\forall h_i \in H_k(t)} \left(c_i^t + a_i^{t+1} - a_i^t\right)\right].$$

Under the above conditions, we can analyze the dynamics of the distributions of income and wealth.

### (*) *Dynamics of Factor Income Distributions*

In the real economy, an obvious fact is, as a stock variable that can be accumulated, wealth (i.e. valued capital——equities of firms) can be continually transferred from the current generation to the next generation. Correspondingly, labor income is a flow variable, and cannot be accumulated. In fact, labor income of a newborn baby is always zero, and every-one in their lifetime will experience a process of increasing labor income from zero. Based on these facts, and $s_{H_1(t)} > s_{H_j(t)}$ $(\forall H_j(t) \in \mathcal{K}(t),\ j \neq 1, \forall t > 1)$, we have Proposition 2.

Proposition 2: $W_{H_1(t)}$ is defined as the total labor income of the households in $H_1(t)$, and $A_{H_1(t)}$ is their total wealth. Respectively, $W_{H_1(t)}$ and $A_{H_1(t)}$ satisfy

$$\lim_{t \to +\infty} \frac{W_{H_1(t)}}{A_{H_1(t)}} = 0, \quad W_{H_1(t)} \equiv \sum_{\forall h_i \in H_1(t)} \hat{w}_i^t, \quad A_{H_1(t)} \equiv \sum_{\forall h_i \in H_1(t)} a_i^t.$$

Proposition 2 means, for the group of households with the highest saving rate $(H_1(t))$ in economy, the ratio of their total labor income to total wealth will eventually become small enough. Based on this point, we can draw an interesting inference——in the long run $(t \to +\infty)$, what will gradually disappear is the impact of "technical relationships" on the rate of return on market portfolio. Next, we prove this result.

By Proposition 2, it can be known that, when $t \to +\infty$, for the households in $H_1(t)$, we have

$$\frac{\left(C_{H_1(t)} + A_{H_1(t+1)} - A_{H_1(t)}\right)}{A_{H_1(t)}} = R_{t+1} + \frac{D_{H_1(t)}}{A_{H_1(t)}}(R_{t+1} - \bar{r}_{t+1}), \tag{18}$$

where, respectively, $C_{H_1(t)}$ and $D_{H_1(t)}$ are defined by

$$C_{H_1(t)} \equiv \sum_{\forall h_i \in H_1(t)} c_i^t, \qquad D_{H_1(t)} \equiv \sum_{\forall h_i \in H_1(t)} d_i^t.$$

(18) is essentially the same as (8). The only difference between (8) and (18) is that we have removed $W_{H_1(t)}$ based on the result of Proposition 2. We introduce the definition, $\gamma_{H_1(t)} \equiv A_{H_1(t)} / A_t$. Because (i) the shares of firm are equal rights in the profits, and (ii) the portfolio of the households in $H_1(t)$ is market portfolio, combined with (14), we have

$$\frac{A_{H_1(t+1)}}{\gamma_{H_1(t+1)}} = \frac{A_{H_1(t)}}{\gamma_{H_1(t)}}(1 + R_{t+1}). \tag{19}$$

Proposition 2 shows, when $t \to +\infty$, $s_{H_1(t)}$, $A_{H_1(t)}$, and $D_{H_1(t)}$ satisfy

$$\frac{A_{H_1(t+1)} - A_{H_1(t)}}{A_{H_1(t)}} = s_{H_1(t)}\left[R_{t+1} + \frac{D_{H_1(t)}}{A_{H_1(t)}}(R_{t+1} - \overline{r}_{t+1})\right]; \tag{20}$$

$$D_{H_1(t)} = \mu_t A_{H_1(t)}. \tag{21}$$

According to the definition of $\gamma_{H_1(t)}$, combined with (14), (18), (20), and (21), we get

$$\gamma_{H_1(t+1)} = \gamma_{H_1(t)} \frac{1 + s_{H_1(t)}\left[R_{t+1} + \mu_t(R_{t+1} - \overline{r}_{t+1})\right]}{1 + R_{t+1}}. \tag{22}$$

Now, $A_{H_1(t)}, D_{H_1(t)}, R_{t+1}$, and $\gamma_{H_1(t)}$ can be determined by the system $E_2$ that includes (19), (20), (21), and (22), after giving the values of exogenous variables $s_{H_1(t)}$, $\mu_t$, and $\overline{r}_{t+1}$. Obviously, the endogenous rate of return on market portfolio, $R_{t+1}$, is independent of either the marginal principle or the input-output relationships (i.e. production functions).

(*) ***Asymptotic Nature of Wealth Distribution***

In order to analyze the dynamics of wealth distribution, we introduce a new assumption, i.e.

$$1 > s_{H_1(t)} > s_{H_2(t)} > ... > s_{H_{\bar{M}}(t)} > 0, \quad \forall t > 1. \tag{23}$$

From (23), we have proposition 3.

Proposition 3: When $|H(t)| \mapsto +\infty$, $|F(t)| \mapsto +\infty$, and $t \to +\infty$, the wealth distribution of households approximates a power law distribution.

*Proof*: (See Appendix A).

### 2.5. Determination of Other Endogenous Variables

Under the above conditions $(|H(t)| \mapsto +\infty, |F(t)| \mapsto +\infty, t \to +\infty)$, we can determine all endogenous variables in the model (at both the micro and macro level).

#### (*) Endogenous Variables at Micro Level

After $R_{t+1}$ is determined by $E_2$ (including (19), (20), (21), (22)), combined with the exogenous $\bar{\mathbf{r}}^{t+1}$, $\bar{r}_{t+1}$ and $\mathbf{B}^t$, (17) can determined $\mathbf{Z}^{t+1}$.

Now, giving $\mathbf{Z}^{t+1}$, the system $E_3$, including (1), (2), (3), (4), (6), (15), and (16), will determine the endogenous variables at the micro level. They include $\mathbf{r}^{t+1}$, $\hat{\mathbf{C}}^t$, $\hat{\mathbf{A}}^t$, $\mathbf{A}^t$, $\hat{\mathbf{W}}^t$, $\hat{\mathbf{D}}^t$ and $p_x^{t+1}$. Simultaneously, the exogenous variables are $\hat{\mathbf{r}}^{t+1}$, $\mathbf{V}^t$, $\mathbf{M}^t$, $\hat{\mathbf{W}}^t / p_x^t$, $\mathbf{Q}^t$, and $\mathbf{P}^t / p_x^t$.

#### (*) Endogenous Variables at Macro Level

Using the variables at the micro level, we can calculate the macro variables from (7), i.e. $C_t$, $A_t$, $W_t$, $D_t$.

**2.6. *Products' Relative Prices and Firms' Profits***

**(\*) *Determination of Relative Prices***

The above analysis means that $R_{t+1}$ and $\mathbf{Z}^{t+1}$ are neither affected by relative prices nor related to the marginal principle. In a perfectly competitive market, general equilibrium analysis can give us a set of relative prices, ( $\hat{\mathbf{W}}^t / p_x^{t+1}$, $\mathbf{P}^t / p_x^t$, $t = 1,2,...$ ). Unlike traditional general equilibrium frameworks, our model does not rely on the perfect competition assumption. In fact, as long as there exists a mechanism determining relative prices, which can be different from general equilibrium, our model will work properly. Considering the deviation of the real economy from the perfect competition framework, it can be seen as the advantage of our model over traditional general equilibrium analysis.

**(\*) *Firms' Profit and Profit Rate***

In the above model, the profit of the firm $f_i$ is related to its net cash flow $\mathbf{m}_i^t$. When we introduce a general equilibrium framework into the above model to determine relative prices, and if all financial dealings (payments, income, etc.) and productions of competitive firms are concentrated in one period, we will get the same results as the traditional models——the profits of competitive firms are zero. Otherwise, the no-arbitrage condition will be violated. Conversely, if financial transactions of firms are distributed in a series of periods, i.e. there is a "cash flow," then the profit rates of firms ($\mathbf{z}^{t+1}$) will be determined by the no-arbitrage principle (see (17)).

If a firm has market power, from a traditional point of view, it can obtain monopoly profit. However, from the perspective of external investors, the rates of returns on the equities of the firms ($\mathbf{z}^{t+1}$), i.e. profit rates on valued capitals, still need to satisfy (17).

## 3. Discussion and Analysis

### 3.1. *Simple Interpretation Based on Intuition*

The neoclassical factor income distribution theory depends on the marginal principle. However, in our analysis, the marginal principle only works at the micro level. Intuitively, because the consideration of profits is based on valued capitals, i.e. the goals of shareholders are to "make money" instead of "make goods," total profits of all firms must be related to money supply in the aggregate sense. This is the essence behind the identity equation (10)——it comes from the aggregation of (3). Starting from this, we can derive the equation system including total assets, the return rate on total assets, debts, wealth and saving rate distributions, etc., and determine rate of return on market portfolio——i.e. the "average" profit rate at the macro level. In this process, the relative prices, determined by technical relationship, no longer affect the "average" profit rate (i.e. $R_{t+1}$) based on total value capital ($A_t$). In fact, (10) and (14), as two identity equations, can simultaneously determine $R_{t+1}$ and $A_t$ without being affected by the technical relationship at the micro level.

Essentially, our model is constructed on two identity equations (i.e. (10), (14)) at the most basic level. In this sense, the new theory is robust. The various factors at the micro level (expectation, decision-making methods, etc.) will not affect the effectiveness of the entire framework.

**3.2.** *Contributions to Cambridge Capital Controversy*

The above analysis can help to solve problems related to the Cambridge capital controversy (CCC) that mainly occurred in the 1950s to 1970s. Due to space limitations, we only discuss three issues: (i) the capital aggregation and factor distribution theory; (ii) reswitching and capital reversing; (iii) Pasinetti paradox.

**(*)** *Capital Aggregation and Factor Distribution Theory*

In 1950s, Joan Robinson's complaints about aggregate production function ignited the CCC that lasted for more than 20 years. J. Robinson (1953-54) correctly pointed out, when there are more than one kind of capital goods, the traditional theory, i.e. profit rate is determined by the marginal contribution of aggregate capital, will inevitably lead to the problem of circular argument. In fact, the theoretical essence of J. Robinson's critique is that, the Wicksell effect makes it impossible to maintain a one-way relationship between value of capital goods and rate of profit (Cohen and Harcourt, 2003).

The main defense of neoclassical economics is based on general equilibrium theory (Lazzarini (2011), Cohen and Harcourt (2003)). In a general equilibrium framework, all markets are simultaneously cleared and the relative prices are determined by the marginal principle. In this process, it is not necessary to introduce aggregate production function. However, this approach is contrary to the unified-rate-of-return theory in classical economics (Garegnani (1976, 1990)), or equivalently, the no-arbitrage principle in modern finance. Generally, in the economy with non-perishable heterogeneous capital goods, the prices of capital goods determined by general equilibrium are often inconsistent with the no-arbitrage principle. In Appendix B, we illustrate this point with a simple example.

In our model, there is a mutually decisive relationship between valued capitals ($\mathbf{A}^t$) and the rates of returns on them ($\mathbf{Z}^{t+1}$). Both of them are independent of capital goods prices and marginal principle. Thus, J. Robinson's complaints (1953-54) have been naturally resolved. In detail, the rate of return on market portfolio ($R_{t+1}$) provides a pricing benchmark for all assets, and in an asymptotic sense, it can be determined by the saving rate of the richest group of household (i.e. $s_{H_1(t)}$, see subsection 2.4). At the micro level, the relative prices of capital goods are determined by technical relationship, and can be calculated by general equilibrium. Since the values of capital goods ($\mathbf{K}^t \mathbf{P}^t$) are independent of $\mathbf{A}^t$ and $A_t$ (i.e. valued capitals), our results above are not surprising.

(*) *Reswitching and Capital-reversing*

Both reswitching and capital-reversing mean that the law of diminishing marginal product may not be satisfied between profit rate and capital-labor ratio (Sraffa (1960), Samuelson (1966)). In fact, reswitching means that it is impossible to order techniques monotonically with rates of profits. In other words, at least in theory, there are no convincing indicators to measure the capital intensities among different techniques. This constitutes a challenge to the mainstream factor income distribution theory. From our point of view, it should be noted that both sides of the CCC, neither Cambridge (UK) nor neoclassical economics distinguishes between valued capital and capital goods (i.e. financial capital and physical capital). Correspondingly, our model introduces both of them simultaneously. In this way, we can solve the problems of reswitching and capital-reversing, or more accurately, make them no longer important.

In our model, factor income distribution at the macro level does not depend on the marginal principle. Any production functions, no matter what form they take, will not affect the factor income distribution at the macro level. Specifically, at the macro level, there is no one-way relationship between valued capitals and return rates on them. More importantly, the analysis of factor income distribution at the macro level does not need to introduce the law of diminishing marginal returns. In other words, regardless of whether reswitching and capital-reversing exist or not at the empirical level, it is neither important nor necessary for the analysis of factor income distribution at the macro level.

Correspondingly, at the micro level, the relative prices of capital goods can be determined by the marginal principle——at least under the assumption of perfect competition. However, this process has no effect on rate of return on market portfolio ($R_{t+1}$). In an asymptotic sense, any relative prices, including relative wages, have nothing to do with $R_{t+1}$ (see subsection 2.4). From an analytical perspective, if reswitching or capital-reversing have a theoretical meaning for analysis, they are only limited to the micro level.

(\*) ***Pasinetti Paradox***

As a critique to neoclassical economics, based on the Kaldor model of distribution (Kaldor, 1955-56), Pasinetti (1962) proposed a theory that rate of profit is determined by the saving rate of the "capitalist class," and not related to the marginal principle. Samuelson and Modigiani (1966) published the theory of opposition as a rebuttal from the neoclassical economics. However, the model of Samuelson and Modigiani (1966) implies an economy without a capitalist class (Pasinetti (1966)). Obviously, it is unrealistic. So, the problem remains unresolved at the theoretical level.

The key to solving this problem is that our model introduces two kinds of capitals——valued capital and capital goods. Our model distinguishes the differences between pricing two concepts of capitals at the macro and micro level. In a sense, our framework absorbs the views of both sides of the debate simultaneously.

At the macro level, subsection 2.4 shows, in an asymptotic sense, $R_{t+1}$ and $\mathbf{z}^{t+1}$ are determined by $s_{H_1(t)}$, and both of them are independent of marginal principle. If the group of households $H_1(t)$ are regarded as a "capitalist class," our results are similar to those of Pasinetti (1962). Correspondingly, neither the model of Pasinetti (1962) nor the neoclassical models distinguish between the above two different concepts of capital.

At the micro level, our model allows the prices of capital goods to be determined by the marginal principle. Thus, our model is consistent with the results of the neoclassical models at this aspect. From a micro-macro perspective, the Pasinetti (1962) model lacks a micro-foundation. Correspondingly, the neoclassical models directly extend the results at the micro level to the macro level, and have led to many unsolved problems (Kirman (1992, 2009)). Our model overcomes these shortcomings.

### 3.3. *General Equilibrium and No Arbitrage Principle*

For the mainstream framework, in a multi-capital-goods economy, there is logical inconsistency between general equilibrium and no arbitrage principle (Garegnani(1976,1990)). From our perspectives, no arbitrage principle is essentially equivalent to unified-rate-of-return theory in classical economics, which can determine the relative ratios among the rates of returns on any portfolios (including individual assets) (see (17)), and the "anchor point" of pricing of (17) is market portfolio rate of return $R_{t+1}$, which is given by the system of equations $E_2$. In contrast, the relative prices of capital goods are determined by general equilibrium, and the values of capital goods ($\mathbf{K}^t \mathbf{P}^t$) are independent of the capital values ($\mathbf{A}^t$ and $A_t$). Therefore, the above inconsistency, between general equilibrium and no-arbitrage principle, will no longer constitute a problem.

In our model, households save the wealth in the value sense (i.e. valued capital, or equities of firms). Thus, if heterogeneous capital goods can be directly saved, then we need to discuss the impacts of the arbitrage behavior of saving capital goods in a period and selling them in another period. In fact, these arbitrage behaviors make general equilibrium and the no-arbitrage principle impossible to maintain intrinsic consistency. In Appendix B, we illustrate this result by a simple example. Correspondingly, the approach of our model is that these arbitrage behaviors are seen as some special "firms" that invest in the special "projects," and investors gain profits by investing in these "firms." In this way, the entire analysis can be integrated into the above framework without affecting our results. Certainly, "uncertainty" is necessary for the model, otherwise the problems will not be solved.

### 3.4. *Uncertainty and More Kinds of Assets*

In the above analysis, it is important to introduce uncertainty and enough heterogeneous households and heterogeneous firms (i.e. $|H(t)| \mapsto +\infty$, $|F(t)| \mapsto +\infty$). If these two conditions are not met, neither (17) nor the results in subsection 2.4 can be maintained. For our model, matter are both uncertainty and the more households, the more kinds of assets. Here, the uncertainty comes from random changes of exogenous parameters.

In a deterministic model, there exists intrinsic inconsistency between no-arbitrage principle and relative prices determined by general equilibrium——unless all products are perishable or homogeneous (see Appendix B). So, to establish a logically consistent profit theory, or to solve theoretical problems related to the CCC, it is necessary to directly construct the model based on the economy with uncertainty, a large number of heterogeneous households, firms, and products (including capital goods). From a micro-macro perspective, our results can be seen as the "emerged" macro-characteristics when the number of individuals increases.

Slightly ironically, in the history of macroeconomic analysis, the simplifications in these aspects did not solve or clarify the problems and instead caused the problems to become complicated and controversial.

### 3.5. *Changes in Product Basket, Households and Firms*

Traditionally, general equilibrium frameworks do not involve changes in households and firms, and hardly have the product-basket-changes entered the mainstream analysis——the model of Stokey (1988) may be an exception, although her model introduces some unconvincing and unrealistic assumptions. Here, the major difficulty comes from the fact that it is difficult for general equilibrium framework to trace the dynamics of equilibriums in a multi-product or multi-capital-goods economy.

Let us first look at the impact of product-basket-changes. From period $t$ to $t+1$, the product basket changes from $Q(t)$ to $Q(t+1)$, and it will change the relative prices in period $t+1$. Thus, in principle, product-basket-changes may have an indirect impact on rate of return on market portfolio by affecting relative wages (see the system of equation $E_1$, in subsection 2.3). However, in an asymptotic sense, this impact will eventually disappear (see subsection 2.4).

Slightly more complicated is the impact of the changes in households and firms. Similar to product-basket-changes, in the long run, what will gradually disappear is the impact of relative-price-changes on rate of return on market portfolio. In detail, changes in households and firms may affect budget constraints of households (see (3)) at the micro level. However, when there exist asset transfer mechanisms among households or among firms, we can modify some of the equations in the system $E_1$ (e.g. (3)), and perform recursive calculations for all subsequent periods. This process does not affect the above results, and there is no any inconvenience. In the real economy, asset transfer mechanisms among households include gifts and inheritance of property, and those among firms are bankruptcy or restructuring mechanisms of firms.

### 3.6. *Heterogeneous Beliefs —— No Impact On Main Results*

In above analysis, (17) is based on the important contribution of Ross (1976). However, it brings a potential problem or "bug." The model of Ross (1976) implicitly requires the expectations of investors are homoge-

neous——i.e. all investors hold the same beliefs in the variance-covariance matrix of the rates of returns on all individual assets. However, it may not be met in the real world. In fact, there may be no equilibrium in the economies with heterogeneous beliefs. After carefully examining the above equations, we can find, even if (17) is no longer satisfied, affected is only the return rates on assets ($\mathbf{z}^{t+1}$) at the micro level, and the other results can still be maintained.

In detail, the system $E_2$ only includes the equations (19), (20), (21) and (22), so it cannot be affected by the failure of (17). At least in an asymptotic sense, our main result——rate of return on market portfolio is independent of marginal productivity——will still be correct. Furthermore, an interesting implication of this result is: even though no arbitrage principle can price the assets at the micro level, the rate of return on market portfolio, i.e. the "average" profit rate at the macro level, is not affected by it.

## 4. Empirical Evidences

### 4.1. *Evidence 1: Test for Market Portfolio Rate of Return*

It is difficult to directly test the equation systems $E_1$ and $E_2$. Fortunately, for the above analysis, there are still testable predictions, and the most important one is the test of formula (8). We next test it using the U.S. data.

In (8), government is ignored. So, we define $G_t$ as the aggregate nominal expenditure of government in period $t$. (8) is transformed into

$$A_{t+1} - A_t = A_t R_{t+1} + W_t + D_t(R_{t+1} - \bar{r}_{t+1}) - G_t - C_t. \tag{24}$$

Let $Y_t \equiv A_t R_{t+1} + W_t$. Rearranging (24), we get

$$R_{t+1} - \bar{r}_{t+1} = \frac{(A_{t+1} - A_t) - s_t(Y_t - G_t)}{D_t s_t}. \tag{25}$$

By introducing the definition equation,

$$g_{t+1} \equiv (Y_{t+1} / Y_t) - 1,$$

(25) can be transformed into

$$R_{t+1} - \bar{r}_{t+1} = \frac{-s_t + s_t(G_t / Y_t) + (A_{t+1} / Y_{t+1})(1 + g_{t+1}) - (A_t / Y_t)}{(D_t / Y_t)s_t}. \tag{26}$$

Similar to most studies, we have to use financial market index as a proxy for the market portfolio. Using the US quarterly data (1964, Q4 ~ 2018, Q4)[5], the validity of equation (26) can be tested. Due to the characteristics of quarterly data, equation (26) is transformed into

$$R_t - \bar{r}_t = \frac{-s_{t-4} + s_{t-4}(G_{t-4} / Y_{t-4}) + (A_t / Y_t)(1 + g_t) - (A_{t-4} / Y_{t-4})}{(D_{t-4} / Y_{t-4})s_{t-4}}. \tag{27}$$

The left side of (27) is the annualized market risk premium.

Using (27), we can calculate the theoretical values of the ex-post market risk premium. Correspondingly, its actual values can be obtained from the S&P 500 index. Respectively, the time series of the theoretical and the actual values are named "NEW_SERIES" and "MKTPT_PREMIUM." The plots of the two series are shown in Figure 1.

---

[5] The data sources can be seen in Appendix C.

[ Insert Figure 1 Here]

In Figure 1, the plots of the two series are very similar and almost overlap, while the correlation coefficient between the two series is 0.749306. The results of the OLS regression are listed below:

$$\text{MKTPT\_PREMIUM} = 0.011596 + 0.471630 * \text{NEW\_SERIRE}$$

$$(\text{s.e. } 0.007412, \text{ t-Stat } 1.560915) \quad (\text{s.e. } 0.028627, \text{ t-Stat } 16.47487)$$

$$(R^2 = 0.561459) \quad (\text{F-statistic} = 271.4212)$$

The above empirical test shows that, we have obtained a simple but accurate formula for ex-post market risk premium. As a powerful evidence, it is a positive support to our model.

### 4.2. *Contributions to Asset Pricing Theory and EPP*

Essentially, the above results (i.e. (26), (27)) provide a pricing formula for rate of return on market portfolio (RRMP). In general, RRMP is often used as the basis for pricing any assets (or portfolios) in asset pricing theories. However, from a logical point of view, RRMP is essentially a "linear combination" of all risk assets in economy, so there is also a "circular argument problem" similar to that in the critique of J. Robinson (1953-54). For the mainstream asset pricing theory, it is a potential loophole at the theoretical level. From a mathematical perspective, the problem here is that for $N+1$ assets (and portfolio) that need to be priced, only $N$ independent equations can be written——unless RRMP can be determined by another equation, so that the number of independent equations increases to $N+1$. In fact, our model solves this problem through substantially in-creasing the number of independent equations by the systems $E_1$ or $E_2$ (see subsections 2.3, 2.4).

A by-product of the above analysis is, our model can perfectly solve the famous theoretical problem in the field of finance——the equity premium puzzle (EPP). The so-called "EPP" is, in general equilibrium framework, the estimated risk aversion coefficient of representative agent is far exceeds realistic and reasonable value, therefore it leads to a "quantitative puzzle" (Mehra and Prescott (1985), Mehra (2003)). The EPP means that, it is difficult for the standard model to match the US data quantitatively. So far, there is no consensus on how to solve the EPP (Mehra (2003)).

From our perspective, the main cause of the EPP is, the results at the micro level are directly extended to the macro level——although this point is implicit in the representative agent framework, which is used by the model of Mehra and Prescott (1985). In more detail, the standard framework (e.g. the model of Mehra and Prescott (1985)) treats RRMP as no different from ordinary assets, and the risk premiums of RRMP and other assets or portfolios are all seen as compensations to risk-averse investors. However, in our analysis, RRMP is a macro variable that represents "average" profit rate in ecnomy, and obeys different rules from those at the micro level. Essentially, RRMP is related to the uniform rate of profit in classical economics. From the equation systems $E_1$ or $E_2$ (see subsection 2.3, 2.4), RRMP in a closed economy has nothing to do with any individual's risk aversion. At the micro level, individual's risk aversions may affect the rates of returns on specific assets. However, at the macro level, RRMP is independent of any investor's risk aversion. In this sense, the cause that leads to the equity premium puzzle must be, the two essentially unrelated variables are put together for discussion and then found to be inconsistent in quantity. In other words, it is a superficial paradox caused by incomplete theories.

### *4.3. Evidence 2: Distributions of Wealth and Income*

The economists acknowledge that wealth of households in economy is subject to Pareto distribution (Gabaix (2016)). In our model, regardless of the initial distributions of income and wealth among households——they are clearly affected by many factors and difficult to analyze, the wealth distribution approximately satisfies the power law in the long run (see Proposition 3, in subsection 2.4). As the Pareto distribution belongs to one of the power law distribution family, our result is consistent with empirical evidences.

In a dynamic sense, the proof process of Proposition 3 implies two testable predictions: (i), the wealth of the economy will gradually concentrate on the richest household groups (see equation (22)); (ii), the distribution of wealth within a richer group of households will be more consistent with power law distribution (or Pareto distribution) than that in a poorer group. More interestingly, these processes are independent of technique relationship. So, our results do not require introducing any assumptions on preferences and technologies. The prediction (i) is consistent with the empirical facts on the wealth distributions of major countries after World War II (see Piketty, Saez (2003), Piketty (2014)). For the prediction (ii), we hope that it can be supported by the more empirical works in the future.

More interesting is another testable prediction of our model. In the proof process of Proposition 3, for the richest group of households in econ-omy, we ignore labor incomes and only consider their property incomes in an asymptotic sense. Thus, from the perspective of empirical testing, there is a new prediction in our model: with the evolution of time, the distribution of wealth in the rich group of households will be closer and closer to the power law. It needs to be tested by future empirical research.

## 5. Summary and Conclusions

Under very general conditions, this paper establishes a micro-macro model for a closed economy. Our contributions include three aspects.

Firstly, we have constructed a logically consistent, complete theory of factor income distribution. New theory solves several traditional theoretical problems——mainly related to the Cambridge capital controversy, including the capital-aggregation problem, reswitching and capital reversing, Pasinetti paradox, intrinsic inconsistency between general equilibrium and the no arbitrage principle.

Secondly, our model provides an approach for establishing a reasonable and logically consistent value theory with micro-foundation. The mainstream economics uses general equilibrium theory as the basic framework for understanding value and price. However, the analysis in this paper shows that, although general equilibrium analysis is reasonable for the determination of relative prices of products (including capital goods) in a perfectly competitive economy, it will lose the effectiveness and lead to many difficult problems when our analysis involves valued capital and profit. Our framework uses the no-arbitrage principle to price valued-capitals (i.e. equities of firms), while the prices of capital goods are deter-mined by traditional analysis. Thus, we unify the two types of value determination methods in an integrated micro-macro framework.

Lastly, we propose a simpler and more reasonable explanation for the equity premium puzzle that has not been solved for more than 30 years.

From an analytical perspective, our model provides a new way to establish a more reasonable macro model with micro-foundation. Most mainstream models "simplify" the analysis and calculation process by introducing representative agent and aggregate production function. It is equivalent to directly extending the results at the micro level to the macro level (e.g. the marginal principle of factor income distribution). These practices have long been proven to be misleading (or mistake) by the contributions of Sonnenschein (1973), Mantel (1974), Debreu (1974), Kirman (1992), etc. Our analytical methods provide a new perspective for solving such problems. Essentially, the model in this paper can be seen as an application of the new framework in the field of factor income distribution. Our model means that even if can work the marginal principle of income distribution at the micro level, it is subject to the different principle at the macro level——unless the model includes only a single product or a single agent.

Our new framework is supported by strong empirical evidences. Furthermore, the assumptions of our model are very simple and general, while the model is logically self-consistent. It increases our confidence in the model. After all, in the final sense, we believe, "The final truth must be simple, beautiful and universal (Shou-Cheng Zhang)."

## Appendix A (Proof of Proposition 3)

We roughly ignore the effect of labor incomes on wealth distribution. From (23) and the Proposition 2, for $\forall \varepsilon_1 > 0$, $|H(t)| \mapsto +\infty$, $t \to \infty$, $\exists x, 1 < x \leq \bar{M}$, we have

$$\left( \sum_{x \leq i \leq \bar{M}} A_{H_i(t)} \bigg/ A(t) \right) < \varepsilon_1, \text{ and } W_{H_i(t)} \big/ A_{H_i(t)} \to 0, \text{ for } \forall i \in [1, x). \quad (A.1)$$

From (A.1), (19), and (20), we get that for $\forall i \in [1, x)$, the group of households $H_i(t)$ satisfies

$$\ln\left(A_{H_i(t)}\right) = \ln\left(A_{H_i(0)}\right) + \sum_{k=0}^{t} \ln\left\{1 + s_{H_i(k)}\left[R_{k+1} + \mu_k(R_{k+1} - \bar{r}_{k+1})\right]\right\}. \quad (A.2)$$

Approximately, linearizing the monotonic relationship in (23), we get

$$s_{H_i(k)} = a_k i + b_k, \quad a_k < 0, \quad b_k > 0. \quad (A.3)$$

Taking a first-order Taylor expansion of (A.2) at $i = 1$, we can get the approximate formula:

$$\ln\left(A_{H_i(t)}\right) = \ln\left(A_{H_i(0)}\right) + \left\{\sum_{k=0}^{t} \frac{a_k\left[R_{k+1} + \mu_k(R_{k+1} - \bar{r}_{k+1})\right]}{1 + (a_k + b_k)\left[R_{k+1} + \mu_k(R_{k+1} - \bar{r}_{k+1})\right]}\right\} \ln(i). \quad (A.3)$$

(A.3) means that we have completed the proof of Proposition 3.

## Appendix B (A Simple Example of Logical Inconsistency)

We consider a closed economy inhabited by a representative agent who lives two periods. In period $t$ ($t = 1, 2$), $x_t$ is the quantity of perishable and unique consumer goods. The representative agent provides 1 unit of labor per period, and her working hours are employed by three firms, whose production functions are $f_1^t$, $f_2^t$, and $f_3^t$, respectively. $f_1^t$, $f_2^t$, and $f_3^t$ satisfy

$$x_t = f_1^t(k_{1,x}^t, k_{2,x}^t, l_1^t); \qquad k_1^t = f_2^t(l_2^t); \qquad k_2^t = f_3^t(l_3^t), \quad t = 1,2 \ . \qquad \text{(B.1)}$$

where $k_{1,x}^t \in R^+$ and $k_{2,x}^t \in R^+$, are the input quantities of the two heterogeneous capital goods of the firm $f_1^t$ in period $t$, respectively. $k_j^t \in R^+$, ($j = 1,2$), are the output quantity of the firm $f_{j+1}^t$, ($j = 1,2$). $l_1^t \in R^+$, $l_2^t \in R^+$, and $l_3^t \in R^+$ are the labor time employed by the firms $f_1^t$, $f_2^t$, and $f_3^t$, respectively. We assume that the production functions $f_1^t$, $f_2^t$, and $f_3^t$ are well-behaved. By definition, we get

$$l_1^t + l_2^t + l_3^t = 1, \qquad t = 1,2 \ . \qquad \text{(B.2)}$$

Let $k_{1,s}^t \in R^+$ and $k_{2,s}^t \in R^+$ be the quantities of two capital goods saved by the representative agent in period $t$. We have

$$k_{j,s}^2 = 0, \quad k_{j,s}^1 + k_j^2 = k_{j,x}^2, \quad j = 1,2 \ . \qquad \text{(B.3)}$$

$$k_{j,x}^1 + k_{j,s}^1 = k_j^1, \quad j = 1,2 \ . \qquad \text{(B.4)}$$

$u(x_1, x_2)$ is the well-behaved utility function of the representative agent. $p_{k1}^t$ and $p_{k2}^t$ are the prices of two capital goods in period $t$, respectively. $w^t$ is the wage rate in period $t$, and $p_x^t$ is the price of consumer goods. We assume that the economy subjected to the "cash-in-advance" constraint. $M_t$ is the money supply in period $t$. The first order conditions are listed below:

$$\frac{1}{p_{k1}^t}\frac{\partial f_1^t}{\partial k_{1,x}^t} = \frac{1}{p_{k2}^t}\frac{\partial f_1^t}{\partial k_{2,x}^t} = \frac{1}{w^t}\frac{\partial f_1^t}{\partial l_1^t} \ , \quad t=1,2\,, \tag{B.5}$$

$$w^t = \frac{\partial f_2^t}{\partial l_2^t} p_{k1}^t = \frac{\partial f_3^t}{\partial l_3^t} p_{k2}^t \ , \quad t=1,2\,, \tag{B.6}$$

$$\frac{1}{p_x^1}\frac{\partial u}{\partial x_1} = \frac{1}{p_{k1}^2}\frac{\partial f_1^2}{\partial k_{1,x}^2}\frac{\partial u}{\partial x_2} = \frac{1}{p_{k2}^2}\frac{\partial f_1^2}{\partial k_{2,x}^2}\frac{\partial u}{\partial x_2} \ . \tag{B.7}$$

The cash-in-advance constraints are

$$p_x^t x_t + p_{k1}^t k_1^t + p_{k2}^t k_2^t = M_t\,, \quad t=1,2 \ . \tag{B.8}$$

On the whole, the system includes a total of 28 independent equations, (B.1~B.8). The endogenous variables include: $p_{k1}^t$, $p_{k2}^t$, $w^t$, $p_x^t$, $x_t$, $l_1^t$, $l_2^t$, $l_3^t$, $k_{1,x}^t$, $k_{2,x}^t$, $k_{1,s}^t$, $k_{2,s}^t$, $k_1^t$, $k_2^t$, $t=1,2$. Thus, general equilibrium can determine all endogenous variables in the economy.

Next, we consider arbitrage-free condition. As our economy is deterministic, the no arbitrage condition requires that the same are the rates of returns on the investment in two capital goods, i.e.,

$$\frac{p_{k1}^2}{p_{k1}^1} = \frac{p_{k2}^2}{p_{k2}^1} \ . \tag{B.9}$$

(B.9) means that the relative prices of the two capital goods remain unchanged during the two periods. However, there is no reason to believe it in the economy. In other words, there exists a logical inconsistency between general equilibrium and no arbitrage principle in our example.

If all capital goods were perishable, or if there were only one type of capital goods, the above problems should no longer appear. In fact, this is the approach of the mainstream frameworks.

## Appendix C (Data Sources and Processing Method)

$A_t$ : Households and Nonprofit Organizations; Net Worth, Level, Billions of Dollars, Quarterly. Source:https://research.stlouisfed.org/fred2/series/TNWBSHNO#

$DY_t$: All Sectors; Credit Market Instruments; Liability, Level-State and Local Governments, Excluding Employee Retirement Funds; Credit Market Instruments; Liability-Federal Government; Credit Market Instruments; Liability, Level) / Gross Domestic Product, (Bil. of US $/Bil. of $), Quarterly. Source : https://research.stlouisfed.org/fred2/graph/?g=14U3#

$Y_t$: Gross Domestic Product, Billions of Dollars, Quarterly, Seasonally Adjusted. Source: https://research.stlouisfed.org/fred2/series/GDP#

$SP_t$: S&P500 Index, Quarterly  1964Q4~2018Q4. Source: http://finance.yahoo.com.

$\bar{r}_{t+1}$: Moody's Seasoned Aaa Corporate Bond Yield©,
Source：https://research.stlouisfed.org/fred2/series/AAA#

$pc_t$: Personal Consumption Expenditures, Billions of Dollars, Quarterly, Seasonally Adjusted.  Source: https://research.stlouisfed.org/fred2/series/PCEC#

$G_t$: Government Consumption Expenditures & Gross Investment, Billions of Dollars, Seasonally Adjusted. Source: https://research.stlouisfed.org/fred2/series/GCE#

$D_t$: $D_t = DY_t \times Y_t$.  $\qquad s_t: s_t = 1 - (pc_t / Y_t)$.

$g_t: g_t = (Y_t - Y_{t-4}) / Y_{t-4}$.  $\qquad R_t: R_t = R_t^q + R_{t-1}^q + R_{t-2}^q + R_{t-3}^q$.

$R_t^q: R_t^q = (SP_t - SP_{t-1}) / SP_{t-1}$.

## REFERENCES

Anderson, P.W., (1972), "More is Different." Science, Vol.177: 393-396.

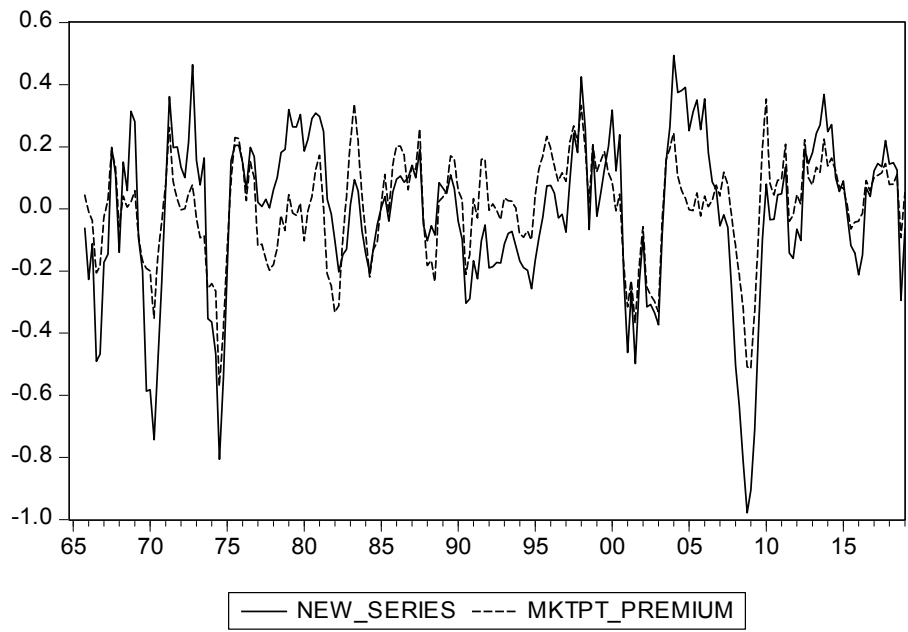

Figure 1. Time Series of Actual and Theoretical Values of $R_{t+1} - \overline{r}_{t+1}$